\RequirePackage{snapshot}       
\documentclass{osa-article}

\journal{oe} 

\usepackage{mathtools}
\usepackage{booktabs}
\usepackage{multirow}
\usepackage[centerlast,font=small,margin=0pt,justification=justified,singlelinecheck=off]{caption}

\newcommand\limitforsmallf{\mathrel{\stackrel{\makebox[0pt]{\mbox{\normalfont\tiny $f \!\!\ll\! f_{\rm FSR}$ }}}{\enspace\longrightarrow\enspace}}}

\begin{document}

\title{Laser Frequency Noise in Next Generation Gravitational-Wave Detectors}

\author{Craig Cahillane,\authormark{1,*} Georgia L. Mansell,\authormark{1,2} and Daniel Sigg,\authormark{1}}

\address{
  \authormark{1}LIGO Hanford Observatory, Richland, WA 99352, USA\\
  \authormark{2}LIGO Laboratory, Massachusetts Institute of Technology, Cambridge, MA 02139, USA
}

\email{\authormark{*}craig.cahillane@ligo.org}



\begin{abstract}
Future ground-based gravitational-wave detectors are slated to detect black hole and neutron star collisions from the entire stellar history of the universe.
To achieve the designed detector sensitivities, frequency noise from the laser source must be reduced below the level achieved in current Advanced LIGO detectors.
This paper reviews the laser frequency noise suppression scheme in Advanced LIGO, and quantifies the noise coupling to the gravitational-wave readout.
The laser frequency noise incident on the current Advanced LIGO detectors is $8 \times 10^{-5}~\mathrm{Hz/\sqrt{Hz}}$ at $1~\mathrm{kHz}$.
Future detectors will require even lower incident frequency noise levels to ensure this technical noise source does not limit sensitivity.
The frequency noise requirement for a gravitational wave detector with arm lengths of 40~km is estimated to be $7 \times 10^{-7}~\mathrm{Hz/\sqrt{Hz}}$.
To reach this goal a new frequency noise suppression scheme is proposed, utilizing two input mode cleaner cavities, and the limits of this scheme are explored.
Using this scheme the frequency noise requirement is met, even in pessimistic noise coupling scenarios.
\end{abstract}



\section{Introduction}
\label{sec:introduction}

Gravitational-wave (GW) astronomy is now a well established field, with regular detections of compact binary coalescences from the ground-based Advanced LIGO and Virgo detectors\cite{GW150914, GW170817, GW190425, GWTC1, GWTC2}. 
Current interferometers are converging on their design sensitivities with incremental upgrades, while the design of a new generation of gravitational-wave detectors is currently under consideration. 
Proposed future ground-based detectors include Cosmic Explorer\cite{Abbott_2017, Hall2021}, a USA-led effort with a Michelson interferometer topology and 40~km long arms, 
and the Einstein Telescope\cite{Punturo_2010}, a European-led effort with triangular interferometer topology and 10~km long arms. 
These next generation detectors, in conjunction with the LISA mission\cite{amaroseoane2017lisa}, seek to answer questions about the universe by observing GW sources out to cosmological distances with extreme precision\cite{reitze2019cosmic}. 
To achieve their proposed sensitivities, headway must be made in understanding and reducing technical noises which currently limit, or lie just below, the noise floor of current detectors.

Laser frequency noise is one technical noise fundamental to laser interferometer GW detectors.
The current Advanced LIGO input optic design is motivated largely by frequency noise suppression, including a 16~m suspended input mode cleaner\cite{Fritschel:2001fx, Kwee:12}.
At GEO600, two 3.9~m input mode cleaners are employed in series, primarily for additional spatial stability\cite{Gossler2003}.

In this paper we present the current understanding of laser frequency noise in Advanced LIGO, which lies within a factor of two of the O3 Advanced LIGO noise floor above a few kHz\cite{Buikema2020}. 
We propose a new laser frequency noise suppression scheme which would reduce this otherwise limiting noise below the proposed Cosmic Explorer sensitivity. 
The proposed scheme utilizes two suspended cavities to mitigate frequency noise incident on the interferometer.

\section{Frequency noise in Advanced LIGO}
\label{sec:frequency_noise_aligo}

In this section we explore how frequency noise affects GW sensitivity in Advanced LIGO.
Here we overview the frequency noise control loops employed, 
the coupling of frequency noise to differential arm motion (DARM),
and the measured Advanced LIGO frequency noise budget.

This information will inform our frequency noise requirements for third-generation interferometers (Section~\ref{sec:requirements}),
and motivate our proposed input optics design which satisfy those requirements (Section~\ref{sec:input_optics_design}).

\subsection{Frequency noise suppression}
\label{subsec:frequency_noise_suppression}

\begin{figure}[t!]
  \includegraphics[width=\linewidth]{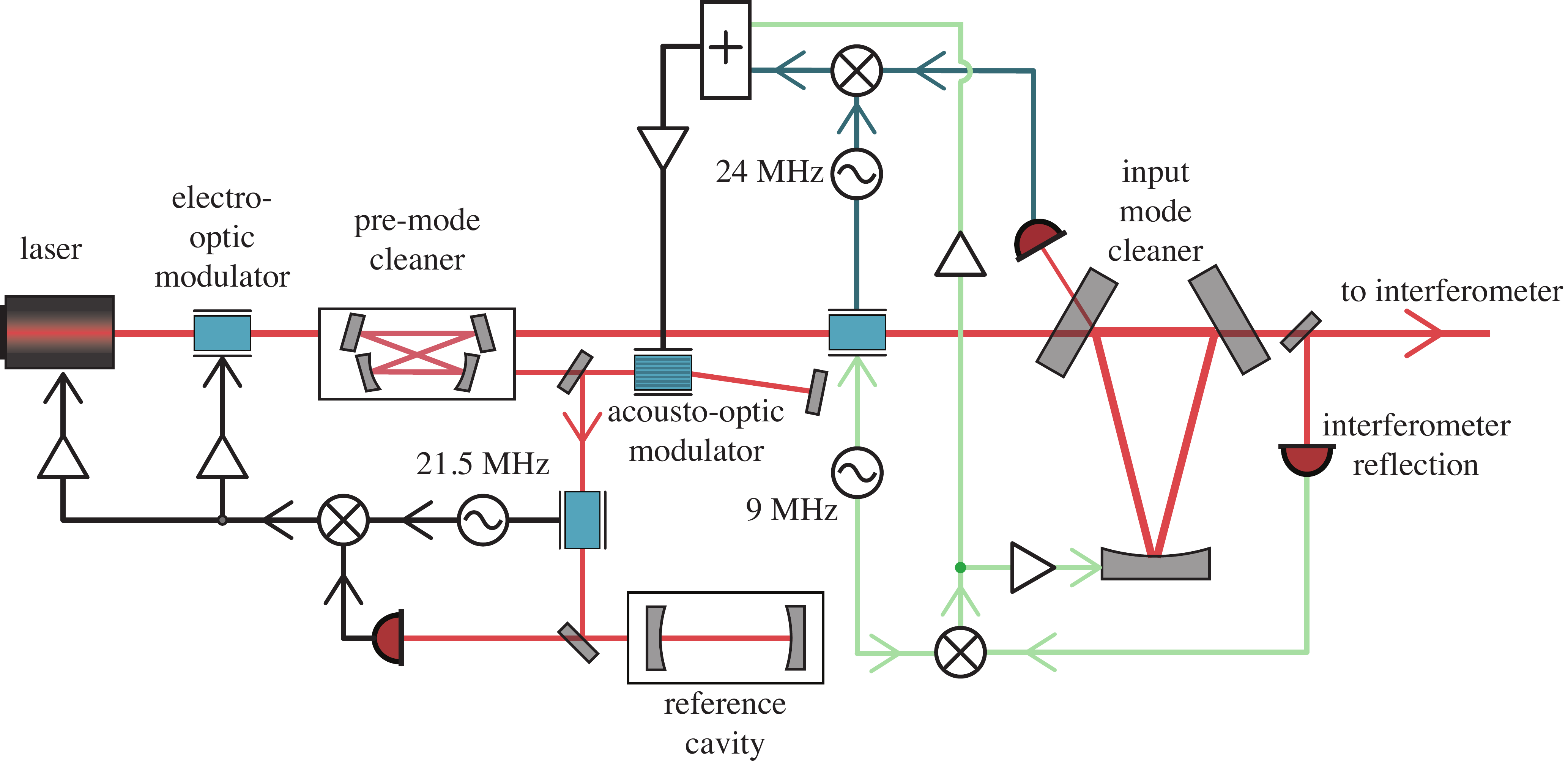}
  \caption{
    Layout of the laser frequency noise suppression system in Advanced LIGO.
    Laser beams are shown in red.
    Electronics for the reference cavity are shown black,
    electronics for the input mode cleaner are dark green,
    and electronics for the common-arm length of the interferometer are mint green.
  }
  \label{fig:aligo_layout}
\end{figure}

Frequency noise must be suppressed to lock the laser to the interferometer, and to avoid frequency noise limiting the GW sensitivity.
The laser frequency noise suppression scheme in Advanced LIGO consists of three main optical elements: 
the reference cavity, the input mode cleaner, and the common arm length. 
Each optical element is locked via the Pound-Drever-Hall (PDH) reflection locking technique \cite{drever:1983}.
The suppression scheme is shown in Fig.~\ref{fig:aligo_layout}.

\subsubsection{Reference cavity}
\label{subsubsec:reference_cavity}

A non-planar ring oscillator (NPRO) laser is amplified to 80 W before passing through a pre-mode cleaner (PMC). 
In transmission of one of the PMC's highly reflective mirrors, a low-power beam double passes an acousto-optic modulator (AOM), acquiring a frequency offset relative to the main interferometer light. 
The frequency-shifted beam then resonates in a 20~cm fixed-spacer reference cavity.
The reflected light contains the reference cavity length signal, which is fed back to the NPRO frequency actuators: the crystal temperature, a piezo-electric transducer (PZT), and a phase correcting Pockels cell (EOM). 
The NPRO laser frequency is locked to the reference cavity length with a 500~kHz bandwidth servo,
achieving a frequency noise level of about $0.01~{\rm Hz}/\sqrt{\rm Hz}$.

\subsubsection{Input mode cleaner}
\label{subsubsec:input_mode_cleaner}

Most of the power transmitted through the PMC passes through an electro-optic modulator (EOM), where phase-modulation RF sidebands are added to the beam. 
One set of RF sidebands is used to lock the input mode cleaner (IMC), while another set is transmitted through the IMC and used to lock the common mode of the main interferometer. 

The beam then propagates to the input mode cleaner, which provides the second stage of frequency noise suppression. 
The laser frequency is locked to the input mode cleaner length at 100~kHz bandwidth, with ``fast'' actuation applied to the AOM on the reference cavity path. 
The performance of the IMC is shown as the blue trace in Fig.~\ref{fig:aligo_freq_noise}, where a frequency noise level of about $1 \times 10^{-4}~{\rm Hz}/\sqrt{\rm Hz}$ is achieved.

\subsubsection{Common-arm cavity length}
\label{subsubsec:common_arm_cavity_length}

The common-arm (CARM) coupled cavity is formed by the arm cavities and the power-recycling mirror in the main interferometer.
The CARM length is the average length of the two arms $L = (L_x + L_y)/2$.
CARM is detected via the reflection from the interferometer.
The CARM length signal forms the final stage of laser frequency actuation with a servo bandwidth of 15~kHz.

There are two feedback paths in the CARM loop, the ``slow'' IMC length path and the ``fast'' laser frequency path.
Below 100~Hz the CARM feedback goes to an input mode cleaner optic suspension.
This forms the ``slow'' CARM feedback loop, which indirectly actuates the laser frequency via the input mode cleaner servo.
This ``slow'' feedback allows the laser light for the interferometer to pass through the IMC as well.

At frequencies above 100~Hz, the CARM feedback is summed with the input mode cleaner error signal.
This forms the ``fast'' CARM feedback loop, which actuates the laser frequency directly.
The function of the ``fast'' loop requires that the RMS frequency correction be less than the IMC bandwidth of 8.8~kHz, in the case of Advanced LIGO.

A cavity is a frequency sensor below the cavity pole, but a phase sensor above.
The CARM double cavity pole $\omega_{\rm cc}$ of the Advanced LIGO interferometer is 0.6~Hz, which makes the reflection detector a phase sensor over the entire frequency band of interest.
This low double cavity pole degrades the interferometer's shot-noise limited sensitivity to frequency noise as the audio frequency $f$ increases.
The performance of the CARM shot noise is shown as the red trace in Fig.~\ref{fig:aligo_freq_noise}, where a phase noise level of about $3.5 \times 10^{-9}~{\rm rad}/\sqrt{\rm Hz}$ is achieved.

\subsection{Frequency noise coupling to DARM}
\label{subsec:frequency_noise_coupling_to_DARM}


An ideal Michelson interferometer with equal arm lengths will reject any frequency noise on the input light away from anti-symmetric port where the GW signal light exits the interferometer.
This property of Michelson interferometer is known as ``common-mode rejection''. 
Advanced LIGO employs a dual recycled Michelson interferometer with arm cavities. 
An imbalance in the arm cavity storage times or an imbalance in their reflectivity will weaken the common-mode rejection\cite{Somiya:2006kb, Izumi_2016, Izumi_fresp1, Izumi_fresp2, Izumi_fresp3}. 

The couplings due to interferometer imbalances are propagated by the fundamental optical mode in the interferometer and experience the double cavity pole as well as the extraction through the signal recycling cavity. 
If the coupling is expressed in displacement sensitivity, the latter term cancels, since a differential arm length change has to go through the same signal recycling cavity to reach the photodetector at the antisymmetric port. 
In Advanced LIGO, at frequencies below 20~Hz additional terms arise from radiation pressure noise, but they tend to be easy to suppress.

The above coupling terms can be written as:
\begin{equation}\label{eq:fcoupling}
    \dfrac{\Delta L_{-}}{\Delta \nu}(f) = 
      \dfrac{\lambda}{2 \omega_{\rm c}} \dfrac{1 + r_{\rm a}}{r_{\rm a}'} \dfrac{1}{1 + s_{\rm cc}}
        \left[ 
          - \delta r_{\rm a} 
          - \dfrac{\delta \omega_{\rm c}}{\omega_{\rm c}} (1 + r_{\rm a}) 
          + \dfrac{l_{\rm sch} r_{\rm a} \omega_{\rm c}}{c} \left(1 - \dfrac{s_{\rm c}}{r_{\rm a}}\right)(1 + s_{\rm c}) 
        \right] 
      + k_{\rm HOM}
\end{equation}
where $L_{-} = L_x - L_y$ is the differential arm length,
$\Delta \nu$ is the amplitude of the laser frequency noise,
$\lambda$ is the laser wavelength,
$r_{\rm a}\approx 1$ is the average arm cavity reflectivity,
$r_{\rm a}'$ is the derivative of the arm reflectivity with respect to the round trip phase,
$\omega_{\rm c}$ is the average angular frequency of the arm cavity pole,
$\delta r_{\rm a}$ is the difference in reflectivity between the two arm cavities,
$\delta \omega_{\rm c}$ is the difference in the arm cavity pole frequency,
$l_{\rm sch}$ is the Schnupp asymmetry\cite{Schnupp:1988}, and
$k_{\rm HOM}$ describes the coupling through higher order optical modes.
The frequency dependence is contained in the cavity pole terms, $s_{\rm c}$ and $s_{\rm cc}$, which are defined as
\begin{align}\label{eq:cavity_poles}
  s_{\rm c} &= i \dfrac{\omega}{\omega_{\rm c}},                      &\quad s_{\rm cc} &= i \dfrac{\omega}{\omega_{\rm cc}} \\
  \omega_{\rm c} &= f_{\rm FSR} \log\left(\dfrac{1}{r_i r_e}\right),  &\quad \omega_{\rm cc} &= f_{\rm FSR} \log\left(\dfrac{1 + r_p r_i}{r_i r_e + r_p r_e (r_i^2 + t_i^2)}\right)
\end{align}
where 
$\omega$ is the audio angular frequency,
$\omega_{\rm cc}$ is the double cavity pole,
$r_p$ is the power recycling mirror amplitude reflectivity,
$r_i$ is the input test mass amplitude reflectivity, and 
$r_e$ is the end test mass amplitude reflectivity\cite{Izumi_2016, Izumi_fresp1, Izumi_fresp2, Izumi_fresp3}. 
Parameters for LIGO Hanford during the third observation run are listed in Table~\ref{tab:TF_params} \cite{CahillaneThesis, Buikema2020}.

\begin{table}[ht]
    \centering
    \caption{\label{tab:TF_params}
      Interferometer parameters for Advanced LIGO and Cosmic Explorer. 
      Most values are design parameters with the exception of the rows marked by (*), 
      where the Advanced LIGO parameters were measured with the detector at the LIGO Hanford Observatory, 
      and where the Cosmic Explorer parameters are estimates.
    }
    \footnotesize
    \setlength{\tabcolsep}{10pt}
    \renewcommand{\arraystretch}{1.1}
    \centering
    \begin{tabular}{ l l r r l }
        \toprule
        \multirow{2}{*}{Parameter}&\multirow{2}{*}{Symbol}&\multicolumn{2}{c}{Values}&\multirow{2}{*}{Units}\\\cline{3-4}
        && LIGO & CE & \\
        \midrule
        Arm length                            & $L$                           & 4       & 40        & km    \\
        Arm cavity pole                       & $\omega_{\rm c}/2 \pi$        & 43.9    & 4.2       & Hz    \\
        Arm cavity pole difference$^*$        & $\delta \omega_{\rm c}/2 \pi$ & 2.4     & 0.04-0.2  & Hz    \\
        Double cavity pole                    & $\omega_{\rm cc}/2 \pi$       & 0.6     & 0.06      & Hz    \\
        Arm reflectivity                      & $r_{\rm a}$                   & 0.9989  & 0.9936    & --    \\
        Arm reflectivity difference$^*$       & $\delta r_{\rm a}$            & 31      & <5000     & ppm   \\
        Arm reflectivity derivative wrt phase & $r_{\rm a}'$                  & 272     & 284       & --    \\
        Schnupp asymmetry                     & $l_{\rm sch}$                 & 8       & 1.4       & cm    \\
        HOM coupling$^*$                      & $k_{\rm HOM}$                 & 0.2--1  & 0.2--1    & fm/Hz \\
        Differential readout pole             & $\omega_{\rm rse}/2 \pi$      & 410     & 825       & Hz    \\
        \bottomrule
  \end{tabular}
\end{table}

The coupling through the difference in the arm cavity pole frequency tends to dominate over the difference in reflectivity or the Schnupp asymmetry. 
Experimentally at LIGO Hanford in O3, we found that the difference in the arm cavity poles leads to a good approximation of the laser frequency noise coupling below 1~kHz. 

At higher frequencies a flat term dominates the coupling which is due to higher order optical modes. 
Higher order modes on the input light are not resonant in the CARM cavity and do not experience any frequency dependence due to the double cavity pole. 
Higher order modes can be transformed back to the fundamental mode by a mode mismatch between the two arm cavities. 
A differential mode mismatch will inject this light directly into the signal recycling cavity, where it will experience the same extraction efficiency to the anti-symmetric port as light from a differential Michelson arm length change. 
This light also has the same frequency dependence as a differential arm length change, so this effect will show as an additional flat term, when the coupling is expressed in displacement sensitivity. 
The correct coupling factor is difficult to predict, since it depends on the exact optical mode composition of the input light as well as the modal difference between the two arm cavities. 
Experimentally at LIGO Hanford in O3, we found this term to be between $0.2-1 \times 10^{-15}$~m/Hz, with the exact number depending on the thermal state of the interferometer which we can partially control\cite{Brooks:16}.



\subsection{Frequency noise budget}
\label{subsec:frequency_noise_budget}

\begin{figure}[t]
  \center\includegraphics[width=\linewidth]{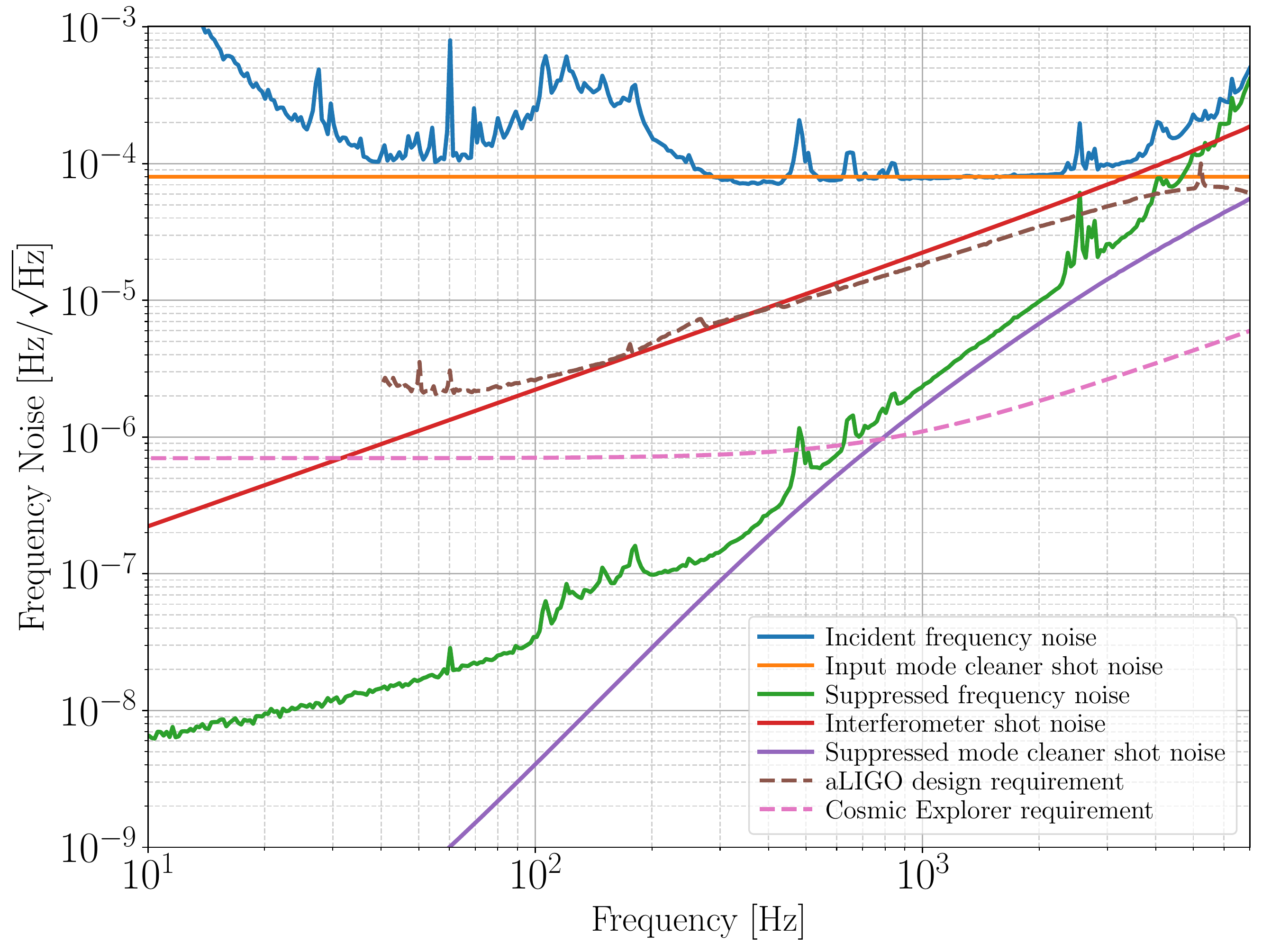}
  \caption{
    The LIGO Hanford observation run three frequency noise budget.
    The measured frequency noise incident on the interferometer, as measured by the common mode control signal, is the blue curve.
    The measured frequency error signal suppressed by the common mode servo is the green curve.
    Red shows the shot noise estimate for the interferometer reflection sensor, while orange shows the shot noise estimate for the input mode cleaner.
    Input mode cleaner shot noise is witnessed by the interferometer as true frequency noise, and is suppressed to the level seen in purple.
    Interferometer reflection shot noise is the ultimate limit for the laser frequency control loop.
    The dashed brown curve depicts the amount of frequency noise that would produce a signal in the gravitational wave readout channel that is equivalent to the Advanced LIGO design sensitivity. 
    The true frequency noise needs to be significantly below this curve to prevent any degradation to the sensitivity.
    The dashed pink curve represents an estimate of the frequency noise requirement in Cosmic Explorer from Eq.~\ref{eq:CEreq}.
    It includes a factor of ten safety margin.
  }
  \label{fig:aligo_freq_noise}
\end{figure}

Fig.~\ref{fig:aligo_freq_noise} shows the frequency noise measured in the Advanced LIGO interferometer \cite{CahillaneThesis}. 
The frequency noise incident on the interferometer is suppressed by the input mode cleaner servo to a level around $8 \times 10^{-5}~{\rm Hz}/{\sqrt{\rm Hz}}$, limited by the shot noise of the input mode cleaner photodetector. 
The common mode servo suppresses the frequency noise further to a phase noise level of about $5 \times 10^{-9}~{\rm rad}/{\sqrt{\rm Hz}}$, limited by shot noise on the interferometer reflection photodetector.

Fig.~\ref{fig:aligo_freq_noise} shows a curve representing the frequency noise that corresponds to the Advanced LIGO design sensitivity. 
It is based on the ratio of the measured frequency to DARM transfer function in $\mathrm{m/Hz}$ and the estimated Advanced LIGO DARM amplitude spectral density in $\mathrm{m/\sqrt{Hz}}$. 
In reality, we would like the laser frequency noise to be a good margin below the Advanced LIGO sensitivity, so that it becomes negligible in the gravitational wave readout channel. 
In Advanced LIGO, the design requirement for technical noises, such as laser frequency noise, is to be ten times below the detector sensitivity.

\section{Requirements and Challenges in Next Generation Detectors}
\label{sec:requirements}

One of the main features of next generation gravitational wave observatories is their size. 
Einstein Telescope\cite{abernathy:2011b} is a triangular design with 10~km long arm cavities. 
Cosmic Explorer\cite{reitze2019cosmic} is an L-shaped detector with 40~km long arm cavities. 
The longer arm cavities will lead to better sensitivity to gravitational waves. 

In the case of Cosmic Explorer, the double cavity pole $\omega_{\rm cc}$ will be an order of magnitude lower than Advanced LIGO. 
Looking at the frequency to DARM coupling function in Eq.~\ref{eq:fcoupling}, the longer arms and reduced double cavity pole will cancel out at GW detection band frequencies. 
Hence, the laser frequency noise couplings due to the imbalances of the two arm cavities will stay the same---assuming all other relevant parameters are unchanged. 
The coupling through higher order optical modes is insensitive to the double cavity pole and one might expect the coupling to stay the same as well. 

While one might expect improvements in the optical surface quality for future generation detectors, the test masses have to be significantly larger to accommodate the larger optical beam size. 
For a design like Cosmic Explorer the optical beam size in an arm cavity is expected to be about twice the size of Advanced LIGO\cite{Dwyer:2015PRD, Hall2021}. 
Furthermore, higher circulating power and larger beam sizes will make thermal lensing on optics in CE similar to Advanced LIGO.
Hence we do not expect significant improvements in the optical beam quality for future detectors which employ high laser power and fused silica test masses.
We can estimate a laser frequency noise requirement for Cosmic Explorer assuming the same coupling of $0.2-1 \times 10^{-15}$~m/Hz as Advanced LIGO.

Cosmic Explorer has a projected displacement sensitivity that is about a factor of two lower than Advanced LIGO at high frequencies. 
Therefore, the laser frequency noise requirement will be a factor of two lower than in Advanced LIGO. 
An additional safety factor of ten is incorporated into the CE requirement to ensure that frequency noise does not limit displacement sensitivity.

If we assume the highest measured value $k_{\rm HOM} = 1 \times 10^{-15}~\rm m/Hz$ for the frequency noise coupling through higher order modes, 
we estimate the laser frequency noise requirement of Cosmic Explorer:
\begin{equation}\label{eq:CEreq}
    F_{\rm in}(f) \leq 7 \times 10^{-7} \frac{\rm Hz}{\sqrt{\rm Hz}} \times \left| 1 + \frac {i f}{825 {\rm~Hz}} \right|.
\end{equation}
The requirement in Eq.~\ref{eq:CEreq} is valid for $f > 10$~Hz as long as cavity pole imbalance between the two arm cavities is not larger than 1\%.
Fig.~\ref{fig:aligo_freq_noise} shows the frequency noise requirement for Cosmic Explorer based on the parameters in Table~\ref{tab:TF_params}. 

In Advanced LIGO, a high bandwidth servo loop is actively suppressing frequency noise by taking the Pound-Drever-Hall error signal in reflection of the interferometer and feeding it back to the laser\cite{Fritschel:2001fx}. 
There are two problems with this suppression scheme for third-generation detectors.
First, the sensing noise for the CE reflection photodetector will be quite high for $f > 100$~Hz, which limits the amount of suppression one can achieve. 
Second, the unity gain frequency of this servo is limited to about half the free spectral range of the arm cavity\cite{Malik:2004}. 
In Advanced LIGO, this is about 15~kHz, while for Cosmic Explorer the servo bandwidth would be limited to about 1.5~kHz.

\begin{figure}
  \centering
  \includegraphics[width=\linewidth]{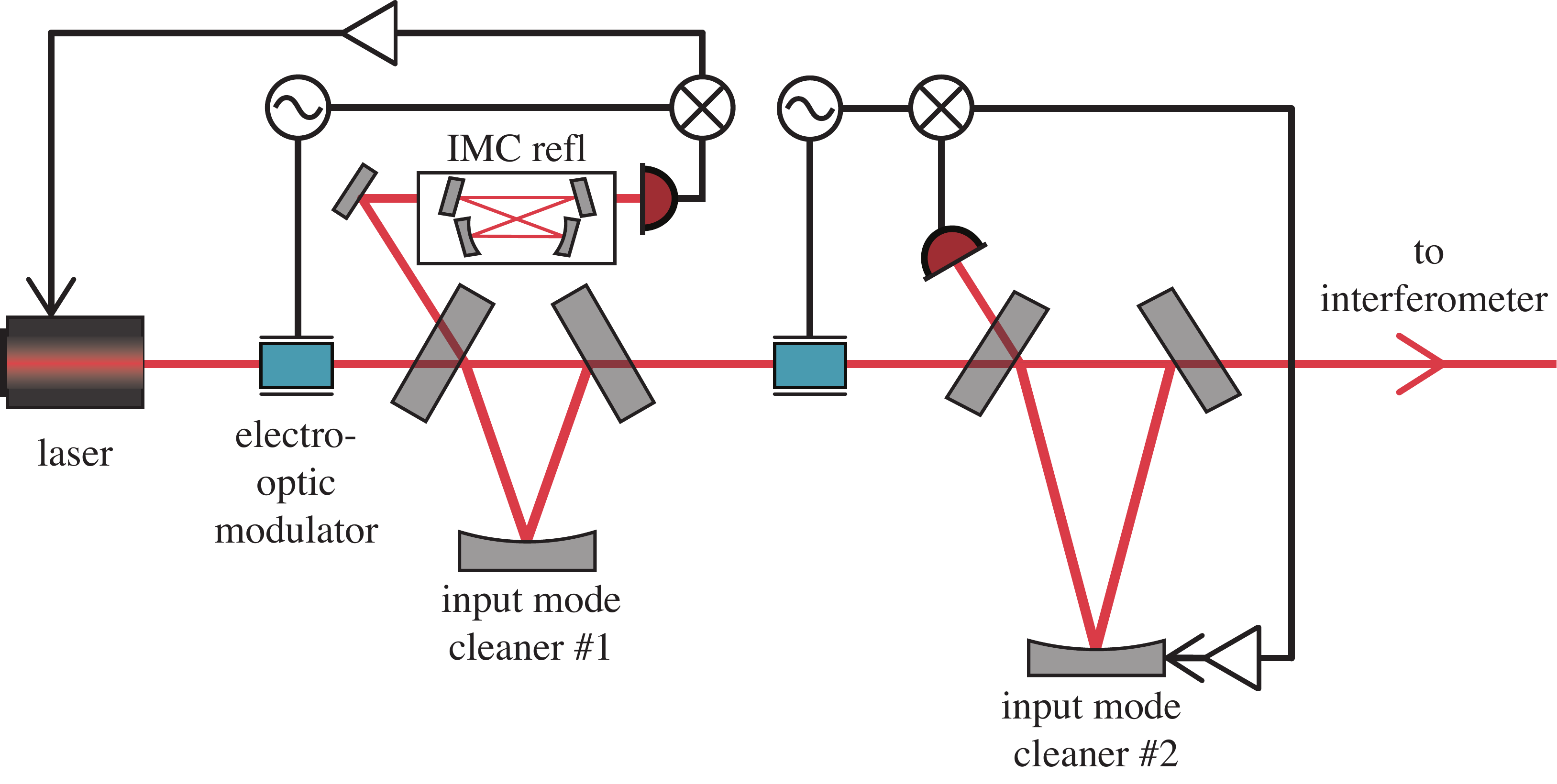}
  \caption{
    Proposed laser frequency noise suppression scheme for 3rd generation gravitational wave detectors, utilizing two suspended input mode cleaners. 
    The pre-stabilized laser consists of a NPRO amplified to 500 W of optical power. 
    This beam acquires phase modulation sidebands via two EOMs. 
    The first input mode cleaner (IMC1) is critically coupled, the reflected beam passes through an additional mode cleaner cavity (IMC refl) to filter out higher order modes. 
    The laser frequency is locked to IMC1 using a high bandwidth servo, the actuator is the PZT inside the laser head. 
    The second mode cleaner (IMC2) longer than IMC1, so the two cavities have different free spectral ranges. 
    The length of IMC2 is locked to the laser frequency, with the feedback applied to one of the optic suspensions.
  }
  \label{fig:ce_layout}
\end{figure}

This requires a significant change in how to suppress frequency noise from the laser. 
Since we can no longer rely on an active servo to clean up the frequency noise in the kHz region, we need to improve the filtering that is provided by the input mode cleaner. 
However, the frequency noise of a laser is very large and cannot be suppressed by a realistic optical filter alone. 
We will still require active suppression with a high bandwidth servo loop. 
Fixed spacer cavities are good references at very low frequencies, 
but are limited in length and suffer from mechanical modes at higher acoustic frequencies that limit their usefulness. 

For future generation detectors we propose two suspended input mode cleaners in series. 
The first input mode cleaner is then used to stabilize the laser frequency with a high bandwidth servo loop, while the second input mode cleaner is locked to the incident light with a low bandwidth servo---providing the necessary filtering at higher frequencies. 
A proposed layout is shown in Fig.~\ref{fig:ce_layout}.

We also need to achieve lower sensing noise in the photodetector in reflection of the interferometer. 
Inspecting Fig.~\ref{fig:aligo_freq_noise} we see that the sensing noise of the Advanced LIGO photodetector mostly meets the Cosmic Explorer requirement below 100~Hz, but falls short at 1~kHz---significantly so with the high coupling for higher order modes. 
To meet the requirement below 100~Hz the arm cavity pole frequencies $\omega_{\rm c}$ of Cosmic Explorer must be matched at the 1\% level. 
This is five times better than the worst case we have seen in Advanced LIGO, but in agreement with the best case.

In Advanced LIGO, the power recycling mirror is not exactly matched to the losses of the arm cavities. 
This leads to a static carrier field in reflection that does not contribute to the signal but is responsible for most of the shot noise. 
Cosmic Explorer might need a set of power recycling mirrors with different reflectivities, so one can choose the proper one to match the losses. 

Another way to improve the signal strength in reflection is to implement a non-resonant RF sideband for the Pound-Drever-Hall reflection locking signal. 
A non-resonant RF sideband will have a reflectivity close to unity and will not suffer from the modulation index strength limitations of resonant sidebands that leak to the antisymmetric port. 
Higher order modes contribute to shot noise, but they can be stripped away with an additional mode cleaner cavity in the reflection path.

\section{Input Optics Design}
\label{sec:input_optics_design}

An input mode cleaner\cite{Mueller:2015} is an optical resonator with equal reflectivity of its input and output coupler to maximize the power throughput. 
For our purpose, the cavity can be parameterized by its length, $L$, and the reflectivity of its input and output couplers, $R$. 
Alternatively, we can characterize the cavity by its free-spectral-range frequency $f_{\rm{FSR}}$, cavity pole frequency $f_{\rm{pole}}$, and finesse $\mathcal{F}$:
\begin{equation}\label{eq:FSR}
    f_{\rm{FSR}} = \frac{c}{2 L}\quad{\rm and}\quad
    f_{\rm{pole}} = \frac{f_{\rm{FSR}}}{2 \mathcal{F}},\quad{\rm with}\quad
    \mathcal{F} = \pi \frac{\sqrt{R}}{1-R}.
\end{equation}
The cavity pole frequency $f_{\rm{pole}}$ is related to the storage time $\tau$ through $f_{\rm{pole}} = 1/2\pi\tau$ and determines the frequency response. 
Input laser noise transmitted through the cavity is filtered above the cavity pole. 
Increasing the finesse or increasing the cavity length will both decrease the cavity pole frequency. 
However, mirror losses in the cavity \cite{Isogai2013} will ultimately limit the finesse, since they degrade the power throughput. 
Fig.~4 in \cite{PhysRevD.88.022002} shows the best achievable mirror losses as function of spot size. 

If we approximate the cavity round-trip losses, $\delta_{\rm RT}$, with a power law:
\begin{equation}\label{eq:rtloss}
    \delta_{\rm RT}(L) \approx \delta_0 \left(\dfrac{L}{1~\mathrm{m}}\right)^{\frac{1}{3}} \quad{\rm and}\quad \delta_0 \approx 10~\mathrm{ppm},
\end{equation}
the total loss in transmission of an input mode cleaner becomes
\begin{equation}\label{eq:totloss}
    \delta_{\rm tot}(L) \approx \frac{\mathcal{F}}{\pi}  \delta_{\rm RT}(L).
\end{equation}

\begin{figure}
  \includegraphics[width=\linewidth]{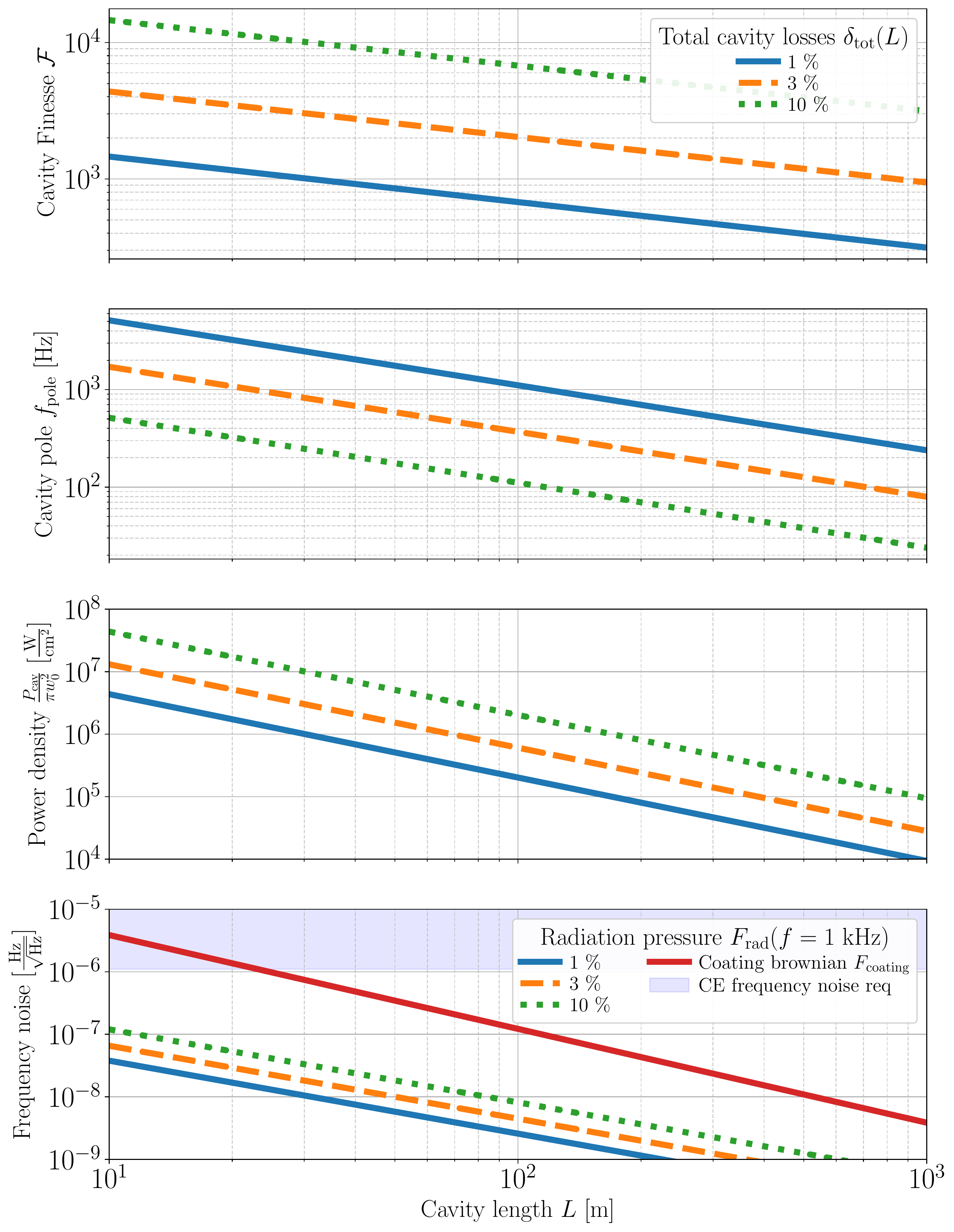}
  \caption{
    Plots of the cavity finesse, pole, maximum power density, and frequency noise at 1~kHz, as a function of cavity length. 
    The blue, orange, and green traces correspond to total transmission losses of 1, 3 and 10\%, respectively, assuming realistic mirror losses. 
    The bottom graph shows the radiation pressure at a fixed frequency of 1~kHz expressed in units of frequency noise assuming 500~W of input power and mirrors that weigh 2.92~kg. 
    It also shows a brown trace representing the frequency noise induced by thermal noise estimated at a frequency of 1~kHz. 
    The shaded region is excluded by the frequency noise requirement of future detectors. 
    The bottom right graph shows the power density on each mirror assuming a cavity with the minimum spot size for the given length. 
  }
  \label{fig:cavity_length_vs_cavity_params}
\end{figure}

Fig.~\ref{fig:cavity_length_vs_cavity_params} shows the cavity pole frequency and the finesse of a cavity, assuming realistic mirror losses, as function of cavity length. 
Since the losses increase with spot size, a longer cavity with its larger beam size ends up with a lower finesse than a shorter one, if we keep the power throughput the same. 
For example, if we tolerate 1\% transmission loss, a cavity of a 100~m length will have a finesse of approximately 700 with a corresponding cavity pole near 1~kHz.

Advanced LIGO is designed for arm cavity powers of 800~kW, which requires about 120~W of light at the input. 
Future detectors want to quadruple the arm powers. 
We assume that they will need around 500~W of input light through the input mode cleaners. 
Frequency noise introduced at the input mode cleaner, including radiation pressure noise, thermal noise, controls noise, must be below the frequency noise requirement.

\subsection{Radiation pressure noise}
\label{subsec:radiation_pressure_noise}

Fluctuations in photon momentum cause random forces on the test masses. 
This radiation pressure noise can be written as:
\begin{equation}\label{radpress}
    F_{\rm rad}(f) = 4 \frac{\sqrt{2 P_{\rm cav} \frac{h c}{\lambda}}}{M L \lambda f^2},
\end{equation}
where $P_{\rm cav}$ is the intra cavity power, 
$\lambda$ the wavelength of the light, and 
$M$ the mass of a mirror. 
We obtain 2.92~kg for a fused silica mirror that is 150~mm in diameter and 75~mm thick. 
Circulating powers will be quite high with $P_{\rm cav}=\frac{\mathcal{F}}{\pi}P_{\rm in}$. 
If we limit the finesse to 3000, we get numbers in the $50-500$~kW range. 
Radiation pressure noise needs to be below the frequency noise requirement above 1~kHz, since there will be no high-bandwidth CARM servo gain to suppress it. 
The bottom left graph of Fig.~\ref{fig:cavity_length_vs_cavity_params} shows the radiation pressure induced frequency noise at 1~kHz as function of cavity length assuming an input power of 500~W. 
It is clearly below the requirement for any reasonable cavity length and we conclude radiation pressure is not a relevant noise source.

The critical power for angular instabilities\cite{Sidles2006} is around 2~MW with mirrors of the above size and mass. 
So, this configuration is stable. We can also calculate the maximum power density on each mirror assuming we choose an optical beam that minimizes the beam size on the mirrors. 
This is shown in the bottom right graphs of Fig.~\ref{fig:cavity_length_vs_cavity_params}. 
The Advanced LIGO input mode cleaner is designed to run at a power density of 350~kW/cm$^2$ or less. 
We have observed laser induced damage on the main test masses in LIGO at much lower power densities, around 5~kW/cm$^2$. 
For future detectors, we want to keep power densities below a MW/cm$^2$ and preferably not higher than in Advanced LIGO.

\subsection{Thermal noise}
\label{subsec:thermal_noise}

Another noise source is thermal noise, which is dominated by coating Brownian noise \cite{NakagawaCTN, ChalermsongsakCTN}. 
We can write:
\begin{equation}\label{eq:coatingthermal}
    F_{\rm coating}(f) = \frac{\sqrt{2} c}{L \lambda}\sqrt{\frac{4 k_B T}{\pi^2 f} \frac{(1+\sigma_s)(1-2\sigma_s)}{E_s} \frac{d}{w^2} \phi_c},
\end{equation}
where $\sigma_s = 0.17$ is the Poisson ratio in silica, 
$E_s = 71$~GPa the Young modulus in silica, 
$w$ is the laser beam spot size, 
$d$ is the thickness of the coating, and 
$\phi_c$ is the loss angle. 
If we assume a loss angle of $\phi_c\approx 4\times 10^{-4}$ and a coating thickness of $d\approx 4.5~\mu$m, we can compute the frequency noise induced by the coating Brownian noise at 1~kHz. 
We can see from the bottom left plot of Fig.~\ref{fig:cavity_length_vs_cavity_params} that for the above coating parameters this requires a mode cleaner that is at least 50~m long to meet the frequency noise requirements.

\subsection{Sensing noise}
\label{subsec:sensing_noise}

We typically use the Pound-Drever-Hall technique\cite{drever:1983} to lock a cavity in reflection. 
The Pound-Drever-Hall technique imposes phase modulated RF sidebands onto the input light and detects the reflected light with a photodiode. 
Since the cavity acts as an FM-to-AM converter when off resonance, demodulating the photodetector signal at the RF frequency yields an error signal which is proportional to the laser frequency deviation, $F(f)$. 
For $f\ll f_{\rm FSR}$, the signal of a Pound-Drever-Hall locking signal can be written as
\begin{equation}\label{eq:PDH}
    P_{\rm PDH}(f) = 2\, \Gamma\, P_{\rm in}  \mathcal{F} \,\frac{F(f)}{f_{\rm{FSR}}} \, \frac{1}{1+i f/f_{\rm{pole}}},
\end{equation}
where $P_{\rm in}$ is the input power, and $\Gamma$ is the modulation depth of the RF sidebands. 

An ideal input mode cleaner is critically-coupled, with equal input and output coupler reflectivities $R$. 
100\% of incident carrier power is transmitted through the input mode cleaner, with no carrier light in reflection. 

In reality, cavity losses will lead to a small amount of reflected carrier light. 
In Advanced LIGO, the reflected carrier light amounts to about $\epsilon=1-2\%$ of the input light. 
The shot noise limited sensitivity is then given by:
\begin{equation}\label{eq:PDHsens}
    S_{\rm PDH}(f) = \sqrt{\frac{2 h c} {\lambda P_{\rm in}}} \sqrt{ \frac32 + \epsilon\frac{2}{\Gamma^2}} \,\left| f_{\rm{pole}} + i f \right|.
\end{equation}

At input powers of tens or hundreds of watts, the amount of light incident on the photodetector becomes a problem. 
If we adjust the modulation depth so that the RF sideband power is roughly equal to the carrier power in reflection, we get $\epsilon\frac{2}{\Gamma^2}\approx 1$. 
If we then further assume that future generation input mode cleaners will be no worse than the current generation, i.e., $\epsilon \lesssim 2\%$, 
the power on the photodetector reaches $\lesssim\!20~{\rm mW}$ with 500~mW of input power. 
Most of this excess carrier power in reflection is due to higher order optical modes on the input light that are rejected by the cavity. 

The actual total power on the photodetector must remain below around $50~\rm mW$; we have assumed $25~\rm mW$ for the sensing noise projections in Fig.~\ref{fig:future_freq_noise_budgets}.
If we want significantly higher input power $P_{\rm in}$ without overloading the photodetector, we will need an output mode cleaner in the reflected beam path to strip away the higher order modes. 
Without it, any increase of the input power requires an optical attenuator in the reflected beam path, which destroys any increase in shot-noise limited sensitivity we may have enjoyed. 
Based on these assumptions we write the following estimates:
\begin{align}\label{eq:IMCsens}
    S_{\rm PDH}(f)    &\approx \left\{ 
      \begin{array}{ll}
        1\times 10^{-6} \dfrac{\rm Hz}{\sqrt{\rm Hz}} \times \sqrt{\dfrac{1\,\rm W}{P_{\rm in}^{\,\rm eff.}}} \times \dfrac{f_{\rm{pole}}}{1\,{\rm kHz}} &{\rm for}\quad f<f_{\rm{pole}} \\
        1\times 10^{-9} \dfrac{\rm rad}{\sqrt{\rm Hz}} \times \sqrt{\dfrac{1\,\rm W}{P_{\rm in}^{\,\rm eff.}}} &{\rm for}\quad f>f_{\rm{pole}}
      \end{array}
    \right.
\end{align}
where $P_{\rm in}^{\,\rm eff.}$ is the effective input power of the $\rm TEM_{00}$ carrier light incident on the mode cleaner which makes it through the reflection output mode cleaner.
We point out that the Cosmic Explorer requirement from Eq.~\ref{eq:CEreq} will require $P_{\rm in}^{\,\rm eff.}\approx 10$~W. 
Fig.~\ref{fig:ce_layout} shows an example layout with a bowtie output mode cleaner, called IMC refl, in reflection of the first IMC.
 


\section{Controls and Performance Projections}
\label{sec:controls}

The frequency noise in transmission of a mode cleaner cavity can be written as function of the incoming laser frequency noise, $F_{\rm in}(f)$, 
the mirror displacement noise, $F_{\rm disp}(f) = F_{\rm rad}(f) + F_{\rm coat}(f)$, 
the sensing noise of the Pound-Drever-Hall reflection locking signal, $F_{\rm sens}(f)$, 
and the feedback compensation network, $G(f)$. 

There are two ways to achieve laser resonance in an optical cavity.
First, we can lock the laser to the cavity by actuating on the laser frequency. 
Second, we can lock the cavity to the laser by actuating on the cavity length. 
The loop gain and bandwidth achievable is vastly different between the two types of locks,
as are the input, displacement, and sensing noise couplings through the mode cleaner.
We refer to Appendix~\ref{sec:appendix_servos} for the full details.

Advanced LIGO uses an NPRO laser as its master oscillator. 
The NPRO free-running frequency noise can be estimated by the following heuristic equation:
\begin{equation}\label{eq:NPROnoise}
    F_{\rm NPRO}(f) =  100~\frac{\rm Hz}{\sqrt{\rm Hz}}\times \frac{100~{\rm Hz}}{f}.
\end{equation}
This is a good approximation in the frequency range from a few Hz up to a MHz. 
In the MHz region excess noise can be observed, but this tends to be irrelevant, since it is high above the observation band. 

The Advanced LIGO laser is first locked to a fixed spacer reference cavity with a bandwidth of 500~kHz. 
This setup provides a frequency noise of $F_{\rm laser}(f) \approx 0.01~{\rm Hz}/\sqrt{\rm Hz}$ in the band from 100~Hz to 10~kHz. 
At lower frequencies, we observe excess noise due to acoustic excitations, whereas at higher frequencies, mechanical modes of the fixed spacer reference cavity can show up prominently. 
There is no real need for reference cavity and one could lock the master oscillator directly to the first input mode cleaner. 
However, this will require a mode cleaner servo with very high gain and also high bandwidth.

For our noise projections, we define a simplified feedback compensation network, $G(f, f_{\rm UGF}, n)$, that is parameterized by the unity gain frequency, $f_{\rm UGF}$, and the number of low frequency boost gain stages, $n$. 
We write
\begin{equation}\label{eq:servo}
    G(f, f_{\rm UGF}, n) = \left( \frac{f_{\rm UGF}}{i f} \right)^{n+1} \left( \frac{1 + \frac{5 i f}{f_{\rm UGF}}}{\left|1+5 i\right|} \right)^{n} \frac{\left|1+\frac{i}{5}\right|}{1+\frac{i f}{5 f_{\rm UGF}}} \left( 1+i \frac{f}{f_{\rm pole}} \right)
\end{equation}
where we added a pole at five times the unity gain frequency for a faster roll off. 
We also added a zero at the cavity pole frequency to compensate for the low pass filtering provided by the optical resonator. 
For long cavities with a free spectral range not far from the unity gain frequency, we will also need notches at the free spectral range frequencies to keep the servo stable. 
The requirement of a bandwidth in the 100~kHz range makes input mode cleaners longer than 300~m impracticable---at least for the first step of laser noise suppression.


\begin{figure}
  \includegraphics[width=\linewidth]{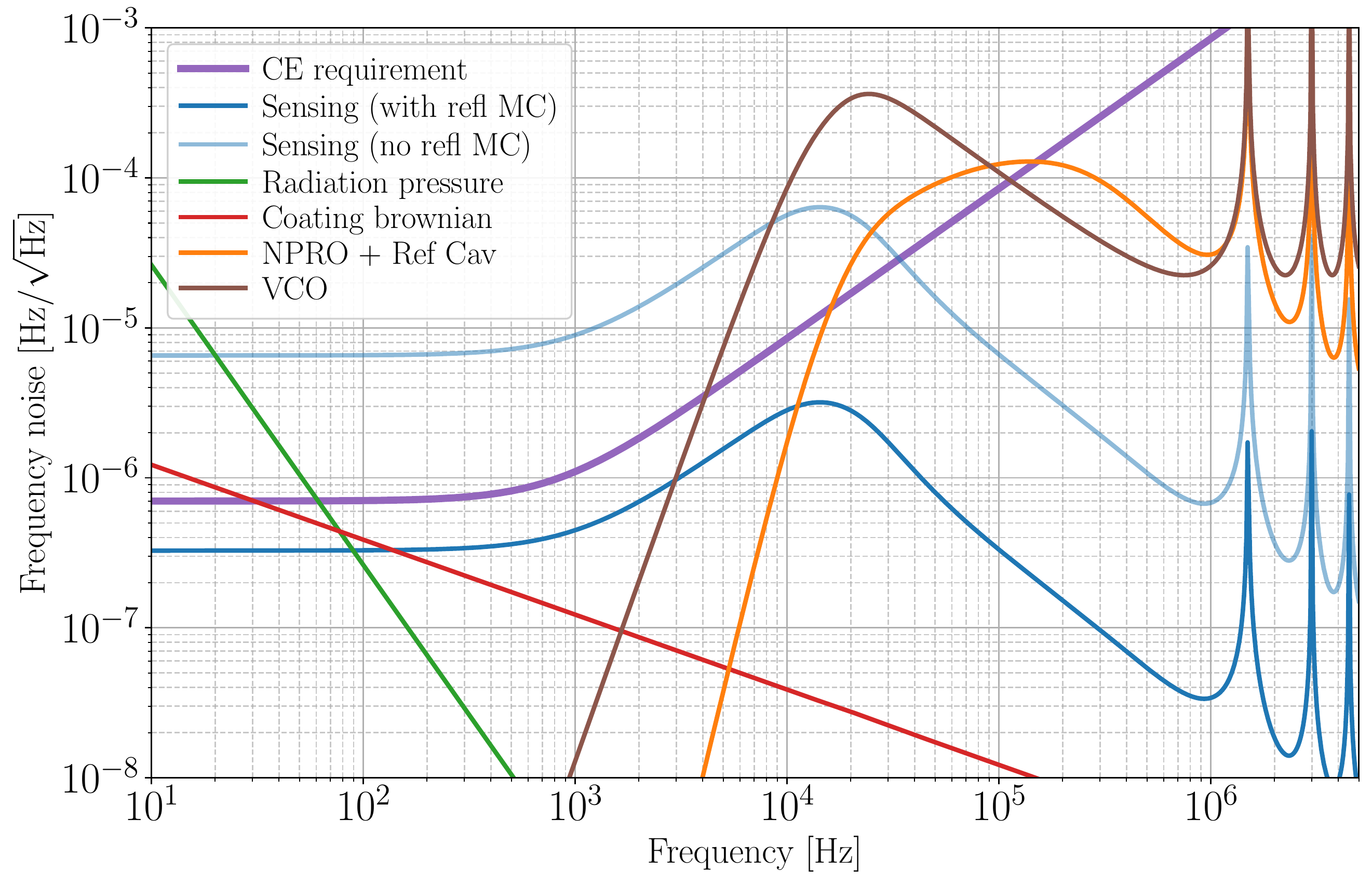}
  \includegraphics[width=\linewidth]{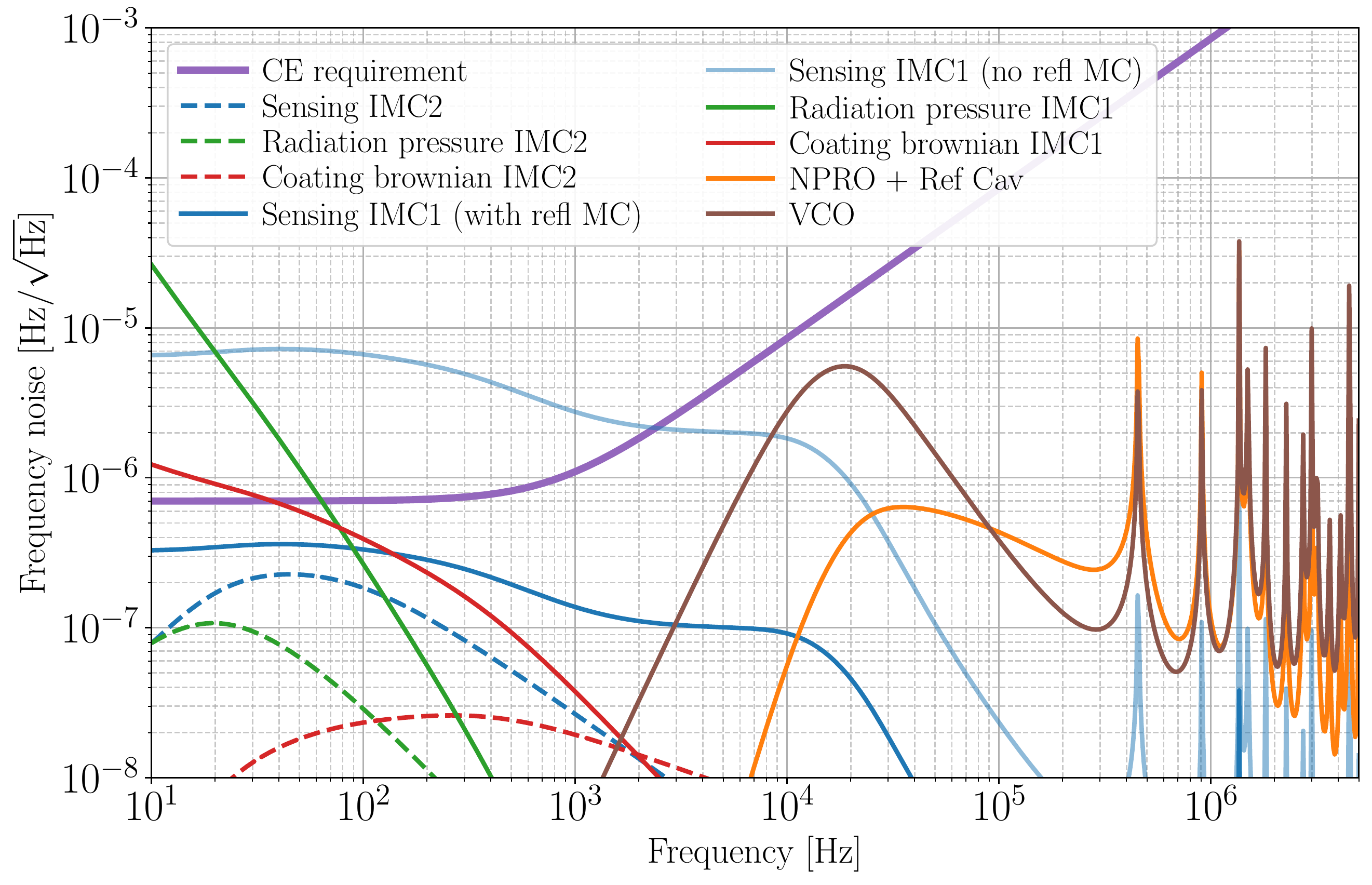}
  \caption{
    Incident frequency noise projections for the proposed input optic configuration with one (top) and two (bottom) input mode cleaners.
    The cavity length of for the first, or only, IMC is 100~m, whereas the second IMC has a cavity length of 330~m.
    The finesse is 700 for all IMCs.
    Solid curves are noises associated with the first, or only, IMC, where the dashed curves are noises from the second IMC.
    The purple band represents the frequency noise requirement at the input of the Cosmic Explorer interferometer, as defined in Eq.~\ref{eq:CEreq}.
    Below a few hundred hertz the output frequency noise of the mode cleaner(s) does not meet the requirement. 
    A common mode servo with a low bandwidth of around 200~Hz is required to achieve the frequency noise requirement.
  }
  \label{fig:future_freq_noise_budgets}
\end{figure}

Fig.~\ref{fig:future_freq_noise_budgets} shows the projected frequency noise at the input of the interferometer after passing through one and two input mode cleaners, respectively. 
The length of the first input mode is 100~m and has a free spectral range of 1.5~MHz. 
The second input mode cleaner is 330~m and has a free spectral range of 0.5~MHz.

The laser is locked to the first or sole input mode cleaner with a unity gain frequency of 100~kHz and three boost gain stages. 
The second input mode cleaner is locked to the laser and has a unity gain frequency of 30~Hz with two boost gain stages. 
Below 1~kHz the displacement noise due to thermal noise and radiation pressure from the first or sole input mode cleaner dominates.

Adding a second input mode cleaner greatly reduces noise at very high frequencies. 
The frequency noise projection indicates that for the detection of gravitational waves the second input mode cleaner is not strictly necessary. 
In both instances, a relatively slow feedback path from the main interferometer with a unity gain frequency around 200~Hz will be sufficient to keep the interferometer locked. 

The main interferometer is an inferior frequency sensor above 1~kHz compared to the first input mode cleaner.
Since the double cavity pole of the interferometer is around 0.05~Hz, the interferometer common mode is effectively a phase detector. 
Furthermore, the reflected carrier power is determined by the losses in the arm cavities, which are difficult to predict exactly, as well as optical higher order modes. 
However, the sensing noise of the interferometer keeps improving towards lower frequencies, whereas the sensing noise of the input mode cleaner stays constant. 
If we assume that the input mode cleaner photodetector can be operated at about 20~times less light power attenuation, the two sensing noise curves cross over around 200~Hz.

A second input mode cleaner has many other advantages. 
It will provide enough filtering at RF frequencies to make the pre-mode cleaner unnecessary. 
In Advanced LIGO the pre-mode cleaner is used to cleanup intensity and frequency noise near the RF sidebands that are used by the Pound-Drever-Hall reflection locking technique. 
A second input mode cleaner also provides additional jitter suppression of the input beam which will most likely be required in future detectors. 
Lastly, it simplifies the intensity noise stabilization, since the two input mode cleaners provide double filtering above their cavity pole frequencies, which is around 1~kHz for the case above.

\begin{table}[ht]
  \centering
  \caption{\label{tab:proposed_ce_controls}
    Proposed Cosmic Explorer input optics controls design parameters.
    The following parameters are used to estimate the CE frequency noise budgets in Fig.~\ref{fig:future_freq_noise_budgets}.
    The reference cavity feedback is optional for the final frequency noise suppression scheme, as the bandwidth of the IMC1 loop can be increased to 500~kHz to achieve the required frequency noise suppression.
    The common-arm feedback goes directly to the laser, but must adjust both IMC lengths such that they accept the laser frequency required for the interferometer, as in Advanced LIGO.
  }
  \renewcommand{\arraystretch}{1.1}
  \centering
  \begin{tabular}{ l l l l }
      Cavity                                & Length                        & Bandwidth & Feedback \\ \hline
      Reference cavity                      & 20~cm                         & 500~kHz   & Laser (optional) \\
      Input mode cleaner 1                  & 100~m                         & 100~kHz   & Laser \\
      Input mode cleaner 2                  & 330~m                         & 30~Hz     & IMC2 suspension \\
      Common-arm cavity                     & 40~km                         & 200~Hz    & Laser and both IMC suspensions \\
\end{tabular}
\end{table}

\section{Conclusions}
\label{sec:conclusions}

We have shown that future gravitational wave detectors such as Cosmic Explorer require a different approach to laser frequency stabilization compared to current detectors. 
The much lower free spectral range of the arm cavities will make it impossible to use a fast laser servo loop from the common mode of the main interferometer to meet the frequency noise requirements. 
Instead, future detectors will have to implement one or two input mode cleaners with a length in the range of 50~m to 300~m to suppress and filter the incoming laser frequency noise. 
A single input mode cleaner of 100~m length can meet the frequency noise requirement of future detectors. 
However, a second input mode cleaner of 330~m length has many advantages in providing additional filtering of the input light, in particular for beam jitter and intensity noise.

\section*{Acknowledgments}
\label{sec:acknowledgments}

LIGO was constructed by the California Institute of Technology and Massachusetts Institute of Technology with funding from the National Science Foundation, and operates under Cooperative Agreement No. PHY-0757058. 
Advanced LIGO was built under Grant No. PHY-0823459. 
The authors also gratefully acknowledge the support of the LIGO Scientific Collaboration Fellows program.

\bigskip

\bibliography{refs}

\begin{thebibliography}{10}
\newcommand{\enquote}[1]{``#1''}

\bibitem{GW150914}
{LIGO Scientific Collaboration} and {Virgo Collaboration}, \enquote{{GW150914}:
  The {Advanced LIGO} detectors in the era of first discoveries,}
  {\protect\JournalTitle{Phys. Rev. Lett.}} \textbf{116}, 131103 (2016).

\bibitem{GW170817}
{LIGO Scientific Collaboration} and {Virgo Collaboration}, \enquote{{GW170817}:
  Observation of gravitational waves from a binary neutron star inspiral,}
  {\protect\JournalTitle{Phys. Rev. Lett.}} \textbf{119}, 161101 (2017).

\bibitem{GW190425}
{LIGO Scientific Collaboration} and {Virgo Collaboration}, \enquote{{GW190425}:
  Observation of a compact binary coalescence with total mass ~3.4 msun,}
  {\protect\JournalTitle{Submitted to ApJ Lett}}  (2020).

\bibitem{GWTC1}
{LIGO Scientific Collaboration} and {Virgo Collaboration}, \enquote{{GWTC-1}: A
  gravitational-wave transient catalog of compact binary mergers observed by
  {LIGO} and {Virgo} during the first and second observing runs,}
  {\protect\JournalTitle{Phys. Rev. X}} \textbf{9}, 031040 (2019).

\bibitem{GWTC2}
{LIGO Scientific Collaboration} and {Virgo Collaboration}, \enquote{{GWTC}-2:
  Compact binary coalescences observed by {LIGO} and {Virgo} during the first
  half of the third observing run,}  (2020).

\bibitem{Abbott_2017}
{LIGO Scientific Collaboration}, \enquote{Exploring the sensitivity of next
  generation gravitational wave detectors,} {\protect\JournalTitle{Classical
  and Quantum Gravity}} \textbf{34}, 044001 (2017).

\bibitem{Hall2021}
E.~D. Hall \emph{et~al.}, \enquote{Gravitational-wave physics with cosmic
  explorer: Limits to low-frequency sensitivity,} {\protect\JournalTitle{Phys.
  Rev. D}} \textbf{103}, 122004 (2021).

\bibitem{Punturo_2010}
M.~Punturo \emph{et~al.}, \enquote{{The Einstein Telescope: A third-generation
  gravitational wave observatory},} {\protect\JournalTitle{Classical and
  Quantum Gravity}} \textbf{27}, 194002 (2010).

\bibitem{amaroseoane2017lisa}
P.~Amaro-Seoane \emph{et~al.}, \enquote{{Laser} {Interferometer} {Space}
  {Antenna},}  (2017).

\bibitem{reitze2019cosmic}
D.~Reitze \emph{et~al.}, \enquote{Cosmic {Explorer}: The {U.S.} {Contribution}
  to {Gravitational-Wave} {Astronomy} beyond {LIGO},}  (2019).

\bibitem{Fritschel:2001fx}
P.~Fritschel \emph{et~al.}, \enquote{Readout and control of a power-recycled
  interferometric gravitational-wave antenna,} {\protect\JournalTitle{Applied
  optics}} \textbf{40}, 4988 (2001).

\bibitem{Kwee:12}
P.~Kwee \emph{et~al.}, \enquote{Stabilized high-power laser system for the
  gravitational wave detector {Advanced LIGO},} {\protect\JournalTitle{Opt.
  Express}} \textbf{20}, 10617--10634 (2012).

\bibitem{Gossler2003}
S.~Go{\ss}ler \emph{et~al.}, \enquote{{Mode-cleaning and injection optics of
  the gravitational-wave detector GEO600},} {\protect\JournalTitle{Review of
  Scientific Instruments}} \textbf{74}, 3787--3795 (2003).

\bibitem{Buikema2020}
A.~Buikema \emph{et~al.}, \enquote{{Sensitivity and performance of the Advanced
  LIGO detectors in the third observing run},} {\protect\JournalTitle{Physical
  Review D}}  (2020).

\bibitem{drever:1983}
R.~Drever \emph{et~al.}, \enquote{Laser phase and frequency stabilization using
  an optical resonator,} {\protect\JournalTitle{Applied Physics B}}
  \textbf{31}, 97--105 (1983).

\bibitem{Somiya:2006kb}
K.~Somiya, Y.~Chen, S.~Kawamura, and N.~Mio, \enquote{Frequency noise and
  intensity noise of next-generation gravitational-wave detectors with {RF/DC}
  readout schemes,} {\protect\JournalTitle{Physical Review D}} \textbf{73},
  122005--17 (2006).

\bibitem{Izumi_2016}
K.~Izumi and D.~Sigg, \enquote{Advanced {LIGO}: length sensing and control in a
  dual recycled interferometric gravitational wave antenna,}
  {\protect\JournalTitle{Classical and Quantum Gravity}} \textbf{34}, 015001
  (2016).

\bibitem{Izumi_fresp1}
K.~Izumi and D.~Sigg, \enquote{Frequency response of the {aLIGO}
  interferometer: part 1,}
  {\protect\JournalTitle{https://dcc.ligo.org/LIGO-T1500325/public}}  (2016).

\bibitem{Izumi_fresp2}
K.~Izumi and D.~Sigg, \enquote{Frequency response of the {aLIGO}
  interferometer: part 2,}
  {\protect\JournalTitle{https://dcc.ligo.org/LIGO-T1500461/public}}  (2016).

\bibitem{Izumi_fresp3}
K.~Izumi and D.~Sigg, \enquote{Frequency response of the {aLIGO}
  interferometer: part 3,}
  {\protect\JournalTitle{https://dcc.ligo.org/LIGO-T1500559/public}}  (2016).

\bibitem{Schnupp:1988}
L.~Schnupp, \enquote{Internal modulation schemes,} in \emph{Presented at the
  European Collaboration Meeting on Interferometric Detection of Gravitational
  Wave, Sorrento, Italy,}  (1988).

\bibitem{CahillaneThesis}
C.~Cahillane, \enquote{Controlling and calibrating interferometric
  gravitational wave detectors,} Ph.D. thesis, California Institute of
  Technology (2021).

\bibitem{Brooks:16}
A.~F. Brooks \emph{et~al.}, \enquote{Overview of advanced ligo adaptive
  optics,} {\protect\JournalTitle{Appl. Opt.}} \textbf{55}, 8256--8265 (2016).

\bibitem{abernathy:2011b}
M.~Abernathy \emph{et~al.}, \enquote{Einstein gravitational-wave telescope
  conceptual design study,} {\protect\JournalTitle{ET technical note,
  ET-0106C-10 (2011)}}  (2011).

\bibitem{Dwyer:2015PRD}
S.~Dwyer \emph{et~al.}, \enquote{Gravitational wave detector with cosmological
  reach,} {\protect\JournalTitle{Phys. Rev. D}} \textbf{91}, 082001 (2015).

\bibitem{Malik:2004}
M.~Rakhmanov, F.~Bondu, O.~Debieu, and R.~L. Savage, \enquote{Characterization
  of the {LIGO} 4 km {F}abry-{P}erot cavities via their high-frequency dynamic
  responses to length and frequency variations,} {\protect\JournalTitle{Class.
  Quantum Grav.}} \textbf{21}, S487--S492 (2004).

\bibitem{Mueller:2015}
C.~L. Mueller \emph{et~al.}, \enquote{The {Advanced LIGO} input optics,}
  {\protect\JournalTitle{Rev. Sci. Instrum.}} \textbf{87}, 014502 (2016).

\bibitem{Isogai2013}
T.~Isogai, J.~Miller, P.~Kwee, L.~Barsotti, and M.~Evans, \enquote{Loss in
  long-storage-time optical cavities,} {\protect\JournalTitle{Opt. Express}}
  \textbf{21}, 30114--30125 (2013).

\bibitem{PhysRevD.88.022002}
M.~Evans, L.~Barsotti, P.~Kwee, J.~Harms, and H.~Miao, \enquote{Realistic
  filter cavities for advanced gravitational wave detectors,}
  {\protect\JournalTitle{Phys. Rev. D}} \textbf{88}, 022002 (2013).

\bibitem{Sidles2006}
J.~A. Sidles and D.~Sigg, \enquote{Optical torques in suspended
  {F}abry-{P}\'erot interferometers,} {\protect\JournalTitle{Physics Letters
  A}} \textbf{354}, 167 -- 172 (2006).

\bibitem{NakagawaCTN}
N.~Nakagawa, A.~Gretarsson, E.~Gustafson, and M.~Fejer, \enquote{Thermal noise
  in half-infinite mirrors with nonuniform loss: A slab of excess loss in a
  half-infinite mirror,} {\protect\JournalTitle{Phys. Rev. D}} \textbf{65}
  (2002).

\bibitem{ChalermsongsakCTN}
T.~Chalermsongsak \emph{et~al.}, \enquote{Broadband measurement of coating
  thermal noise in rigid {F}abry-{P}\'erot cavities,}
  {\protect\JournalTitle{Metrologia}} \textbf{52}, 17--30 (2015).

\bibitem{Bondu:07}
F.~Bondu and O.~Debieu, \enquote{Accurate measurement method of fabry-perot
  cavity parameters via optical transfer function,}
  {\protect\JournalTitle{Appl. Opt.}} \textbf{46}, 2611--2614 (2007).

\end{thebibliography}



\newpage
\appendix
\section{Appendix}
\label{sec:appendix_servos}

\begin{figure}
  \includegraphics[width=\linewidth]{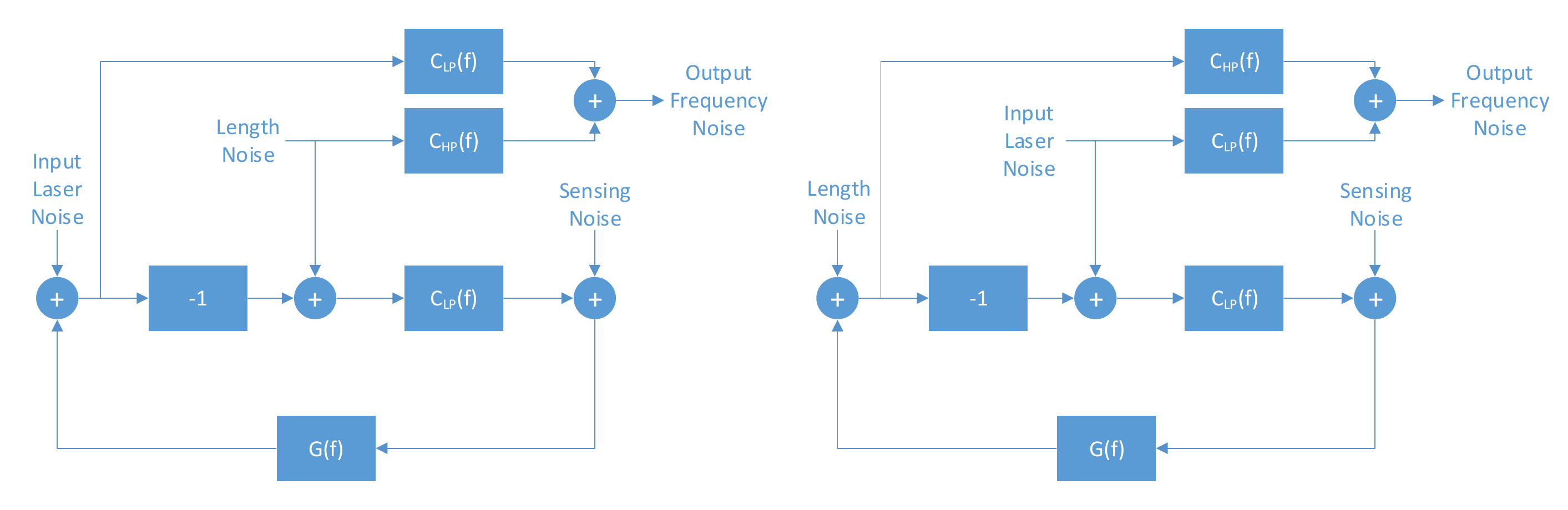}
  \caption{
  The left graph shows the servo diagram for locking a laser to a cavity, whereas the right graph shows the corresponding servo diagram for locking a cavity to the incoming laser light. 
  $C_{\rm LP}(f)$ describes the low pass filtering provided by the cavity, whereas $C_{\rm HP}(f)$ is the corresponding high pass. 
  $G(f)$ is the feedback compensation network designed for locking.
  }
  \label{fig:cavservo}
\end{figure}

Here we overview the two ways to lock a laser to a cavity via Pound-Drever-Hall feedback. 
In Fig.~\ref{fig:cavservo}, the block diagram on the left shows the control loop with feedback going to the laser.
The diagram on the right shows the feedback going to a cavity mirror suspension. 
With $F_{\rm in}(f)$ the incoming laser frequency noise, $F_{\rm disp}(f)$ the mirror displacement noise and $F_{\rm sens}(f)$ the sensing noise, the frequency noise in transmission of the mode cleaner can then be written as:
\begin{equation}\label{eq:IMCtrans}
    F_{\rm trans}(f) =  a_{\rm in}(f) F_{\rm in}(f) + a_{\rm sens}(f) F_{\rm sens}(f)  + a_{\rm disp}(f) F_{\rm disp}(f),
\end{equation}

\noindent For locking the laser to the cavity, we get
\begin{equation}\label{eq:IMCtrans1}
\begin{split}
        a_{\rm in}(f)  &= \frac{C_{\rm LP}(f)}{1+C_{\rm LP}(f) G(f)}, \quad\quad\quad a_{\rm sens}(f) = \frac{C_{\rm LP}(f)G(f)}{1+C_{\rm LP}(f) G(f)}, \\
        a_{\rm disp}(f)&= \frac{C_{\rm LP}(f) G(f)}{1+C_{\rm LP}(f) G(f)} C_{\rm LP}(f) + C_{\rm HP}(f),
\end{split}
\end{equation}
and for locking the cavity to the laser, we get
\begin{equation}\label{eq:IMCtrans2}
\begin{split}
        \overline{a}_{\rm in}(f)  &= \frac{C_{\rm LP}(f)(1+G(f))}{1+C_{\rm LP}(f) G(f)}, \quad\,\, \quad \overline{a}_{\rm sens}(f) = \frac{C_{\rm HP}(f)G(f)}{1+C_{\rm LP}(f) G(f)}, \\
        \overline{a}_{\rm disp}(f)&= \frac{C_{\rm HP}(f)}{1+C_{\rm LP}(f) G(f)},
\end{split}
\end{equation}
where
\begin{equation}\label{eq:IMCtran2}
\begin{split}
        C_{\rm LP}(f)  &= \frac{1-R}{1-R e^{-2\pi i \frac{f}{f_{\rm{FSR}}}}} \limitforsmallf \frac{1}{1+i f/f_{\rm pole}} \\
        C_{\rm HP}(f)  &=1-C_{\rm LP}(f) \quad\enspace\,\limitforsmallf \frac{i f/f_{\rm pole}}{1+i f/f_{\rm pole}}, \quad{\rm with} \\
        R &= \left( \frac{\sqrt{4\mathcal{F}^2+\pi^2}-\pi}{2\mathcal{F}} \right)^2\approx1-\frac{\pi}{\mathcal{F}}.
\end{split}
\end{equation}

\noindent The Pound-Drever-Hall reflection locking signal will experience some complex phase dynamics at the free spectral range frequencies\cite{Bondu:07}. 
We can rewrite Equation~\ref{eq:PDH} as
\begin{equation}\label{eq:PDHfull}
    P_{\rm PDH}(f) = 2\, \Gamma\, P_{\rm in}  \mathcal{F} \,\frac{F(f)}{f_{\rm{FSR}}} \, C_{\rm LP}(f).
\end{equation}
This becomes important when the unity gain frequency of the locking servo approaches the free spectral range. 
The feedback compensation network will require a steep cut-off filter or notches to keep the servo stable.

\end{document}